\documentclass[aps,prd,twocolumn,preprintnumbers]{revtex4-2}
\usepackage[colorlinks=true, linkcolor=blue, citecolor=blue, urlcolor=blue]{hyperref}
\usepackage{bm,latexsym,amssymb,amsmath,mathrsfs}
\usepackage{graphicx}
\usepackage{color}
\usepackage{times,setspace}

\begin{document}

\title{Dynamical chirality production in one dimension}
\author{Tomoya Hayata}
\affiliation{Departments of Physics, Keio University, 4-1-1 Hiyoshi, Kanagawa 223-8521, Japan}
\affiliation{RIKEN iTHEMS, RIKEN, Wako 351-0198, Japan}
\author{Katsumasa Nakayama}
\affiliation{RIKEN Center for Computational Science, Kobe 650-0047, Japan}
\author{Arata Yamamoto}
\affiliation{Department of Physics, The University of Tokyo, Tokyo 113-0033, Japan}
\preprint{RIKEN-iTHEMS-Report-23}

\begin{abstract}
We discuss the quantum computation of dynamical chirality production in lattice gauge theory.
Although the chirality of a lattice fermion is complicated in general dimensions, it can be simply formulated on a one-dimensional lattice.
The chiral fermion formalism enables us to extract the physical part of the chirality production that would be interpreted as the chiral anomaly in the continuous theory.
We demonstrate the computation of the $Z_2$ lattice gauge theory on a classical emulator.
\end{abstract}

\maketitle

\section{Introduction}

The chirality of a massless Dirac fermion is conserved at the classical level but varied by quantum effects. 
This was originally found in anomalous decay of a neutral pion to two photons, and is now known as the chiral anomaly~\cite{PhysRev.177.2426,1969NCimA..60...47B}.
The chiral anomaly switches left-handed and right-handed fermions and thus changes the total chirality of fermionic matter or of the vacuum.
Such a microscopic quantum process shows up as macroscopic phenomena due to its topological nature in accelerator experiments~\cite{Kharzeev:2015znc}, in tabletop experiments~\cite{RevModPhys.90.015001}, and in astrophysical systems~\cite{Kamada:2022nyt}.
A variety of dynamical phenomena are predicted there; e.g., the chiral magnetic wave \cite{Kharzeev:2010gd}, the chiral plasma instability \cite{Akamatsu:2013pjd}, the temporal chiral spiral~\cite{Hayata:2013sea}, etc.

The ab-initio study of chiral dynamics requires simulation of time evolution of quantum field theories.
The time evolution cannot be directly simulated by conventional lattice gauge theory since the Monte Carlo sampling fails~\cite{Alexandru:2020wrj}.
A traditional approach is to solve the inverse problem, which has been applied to real-time quantities such as transport coefficients \cite{Nakamura:2004sy,Meyer:2007ic,Meyer:2007dy,Aarts:2007wj,Ding:2010ga,Aarts:2014nba,Amato:2013naa,Astrakhantsev:2015jta,Astrakhantsev:2017nrs,Astrakhantsev:2019zkr} and sphaleron rate \cite{Kotov:2018vyl,Altenkort:2020axj,BarrosoMancha:2022mbj,Bonanno:2023ljc,Bonanno:2023thi}.
Recently, there has been research on applying machine learning techniques to the inverse problem in lattice simulations~\cite{Rothkopf:2022ctl}.
Remarkably, quantum computing will be an alternative approach in the future.
While the quantum simulation of three-dimensional QCD is a far future dream, the simulation of one-dimensional gauge theory is possible even on near-term quantum computers~\cite{Banuls:2019bmf}.
The special case of one-dimensional gauge theory is analytically solvable, so it will be useful as a good benchmark to check the validity of quantum computing.
The general case is analytically unsolvable and numerically intractable for classical computing.
Chiral dynamics in one-dimensional gauge theory thus is one of the nearby targets for quantum computing~\cite{Kharzeev:2020kgc}.

In this paper, we discuss the real-time evolution of chirality in the Hamiltonian lattice gauge theory.
When we study the time evolution of chirality, we must use the lattice fermion with exact chiral symmetry, that is, the so-called chiral fermion, otherwise the time evolution suffers from artificial chiral symmetry breaking by lattice discretization.
Fortunately, we can avoid this problem without extra cost in the one-dimensional Hamiltonian lattice gauge theory.
We describe the lattice formulation of chirality in Sec.~\ref{sec2} and the time evolution of chirality in Sec.~\ref{sec3}.
The future goal is to simulate the time evolution on a quantum computer and to study the role of the chiral anomaly, but this has not been done yet. 
For the purpose of demonstration, we adopt the $Z_2$ lattice gauge theory as a toy model and show the emulated results in Sec.~\ref{sec4}.

\section{One-dimensional lattice fermion}\label{sec2}

In continuous theory, a massless Dirac fermion has definite chirality.
In other words, a chiral charge is a conserved quantity at least in the classical level.
On a lattice, however, the chiral charge conservation is explicitly violated by the lattice discretization a la Wilson \cite{PhysRevD.10.2445}.
This problem must be cured by changing the formulation of a lattice fermion.
The overlap fermion is a general formulation available in arbitrary dimensions \cite{Creutz:2001wp,Matsui:2005uh,Hayata:2023zuk}, but it is known to be computationally demanding.
In one-dimensional space, there is a much simpler formulation, which was pointed out for the Wilson fermion \cite{Horvath:1998gq}.
The conserved chiral charge can be constructed without changing the Hamiltonian.
Here we summarize the chiral charges of one-dimensional lattice fermions.
We use the lattice unit and eliminate the lattice spacing $a$ throughout this paper.

The Wilson fermion is a two-component spinor $\hat{\psi}(x)= \{ \hat{\psi}_1(x),\hat{\psi}_2(x) \}$.
The gamma matrices $\{\gamma^0,\gamma^1,\gamma^5\}$ are the Pauli matrices.
The massless Hamiltonian is given by
\begin{equation}
\begin{split}
 \hat{H}_f =& \sum_x \bigg\{ r \hat{\psi}^\dagger(x) \gamma^0 \hat{\psi}(x) \\
&-\frac{1}{2} \hat{\psi}^\dagger(x) \gamma^0 (r-\gamma^1) \hat{U}(x) \hat{\psi}(x+1) \\
&- \frac{1}{2} \hat{\psi}^\dagger(x+1) \gamma^0 (r+\gamma^1) \hat{U}^\dagger(x) \hat{\psi}(x) \bigg\} ,
\end{split}
\end{equation}
where $\hat{U}(x)$ is the link operator of a U($1$) gauge field.
We define three kinds of charge operators: the fermion number operator
\begin{equation}
 \hat{Q} = \sum_x \left\{ \hat{\psi}^\dagger(x) \hat{\psi}(x) -1\right\},
\end{equation}
the naive chiral charge operator
\begin{equation}
 \hat{Q}^5_{\rm naive} = \sum_x \hat{\psi}^\dagger(x) \gamma^5 \hat{\psi}(x),
\end{equation}
and the conserved chiral charge operator
\begin{equation}
\begin{split}
 \hat{Q}^5 =& \sum_x \bigg\{ \frac{1}{2} \hat{\psi}^\dagger(x) \gamma^5 \hat{\psi}(x) \\
&+ \frac{1}{4} \hat{\psi}^\dagger(x) \gamma^5 (1-\gamma^1) \hat{U}(x) \hat{\psi}(x+1) \\
&+ \frac{1}{4} \hat{\psi}^\dagger(x+1) \gamma^5 (1+\gamma^1) \hat{U}^\dagger(x) \hat{\psi}(x) \bigg\}.
\end{split}
\label{eqQ5W}
\end{equation}
Taking the basis $\gamma^0=\sigma_1$, $\gamma^1=\sigma_3$, $\gamma^5=\sigma_2$, and $r=1$, we can simplify these operators as
\begin{align}
\begin{split}
 \hat{H}_f =& \sum_x \bigg\{ \hat{\psi}^\dagger_1(x) \hat{\psi}_2(x) + \hat{\psi}^\dagger_2(x) \hat{\psi}_1(x) \\
&- \hat{\psi}^\dagger_1(x) \hat{U}(x) \hat{\psi}_2(x+1) \\
&- \hat{\psi}^\dagger_2(x+1) \hat{U}^\dagger(x) \hat{\psi}_1(x) \bigg\} ,
\end{split}
\label{eqHW2}\\
\hat{Q} =& \sum_x \left\{ \hat{\psi}^\dagger_1(x) \hat{\psi}_1(x) + \hat{\psi}^\dagger_2(x) \hat{\psi}_2(x) -1 \right\} ,
\\
 \hat{Q}^5_{\rm naive} =& \sum_x i \left\{ - \hat{\psi}^\dagger_1(x) \hat{\psi}_2(x) + \hat{\psi}^\dagger_2(x) \hat{\psi}_1(x) \right\} ,
\\
\begin{split}
 \hat{Q}^5 =& \sum_x \frac{i}{2} \bigg\{ - \hat{\psi}^\dagger_1(x) \hat{\psi}_2(x) + \hat{\psi}^\dagger_2(x) \hat{\psi}_1(x) \\
&- \hat{\psi}^\dagger_1(x) \hat{U}(x) \hat{\psi}_2(x+1) \\
&+ \hat{\psi}^\dagger_2(x+1) \hat{U}^\dagger(x) \hat{\psi}_1(x) \bigg\} .
\end{split}
\label{eqQ5W2}
\end{align}
This is a convenient choice for quantum computation because the Hamiltonian is encoded on a one-dimensional chain of qubits \cite{Zache:2018jbt}.
The fermion number operator commutes with the Hamiltonian,
\begin{equation}
 [\hat{H}_f,\hat{Q}] = 0 .
\end{equation}
The two chiral charge operators seem almost the same.
They have the same continuum limit because the difference of these operators is $O(a)$.
The commutation property is however distinct.
The conserved chiral charge operator commutes with the Hamiltonian,
\begin{equation}
\label{eqHQ5W}
     [\hat{H}_f,\hat{Q}^5] =0,
\end{equation}
while the naive chiral charge operator does not,
\begin{equation}
\label{eqHQ5Wnaive}
 [\hat{H}_f,\hat{Q}^5_{\rm naive}] \neq 0.
\end{equation}
Note that the Wilson parameter must be $r=1$ otherwise the commutation relation \eqref{eqHQ5W} is not satisfied.

Since the Kogut-Susskind fermion is often employed in quantum simulation, we here clarify the relation between the Wilson and Kogut-Susskind fermions.
In the Kogut-Susskind fermion formalism, a two-component spinor is constructed by one-component fermions on even and odd sites \cite{KSfermion_1975}.
The Hamiltonian is
\begin{equation}
\begin{split}
 \hat{H}_f =& - \frac{i}{2} \sum_n \bigg\{ \hat{\chi}^\dagger(n) \hat{U}(n) \hat{\chi}(n+1) \\
 &- \hat{\chi}^\dagger(n+1) \hat{U}^\dagger(n) \hat{\chi}(n) \bigg\} .
\end{split}
\end{equation}
Relabeling the fermions as $\chi(2x-1) = -i \psi_o(x)$ and $\chi(2x) = \hat{U}(2x-1) ^\dagger\psi_e(x)$, we can rewrite the Hamiltonian as 
\begin{equation}
\begin{split}
 \hat{H}_f =& \frac{1}{2} \sum_x \bigg\{ \hat{\psi}_o^\dagger(x)  \hat{\psi}_e(x) + \hat{\psi}_e^\dagger(x)  \hat{\psi}_o(x) \\
&- \hat{\psi}_o^\dagger(x+1) \hat{U}^\dagger(2x) \hat{U}^\dagger(2x-1) \hat{\psi}_e(x) \\
&- \hat{\psi}_e^\dagger(x) \hat{U}(2x-1)\hat{U}(2x) \hat{\psi}_o(x+1)  \bigg\} .
\label{eqHKS2}
\end{split}
\end{equation}
This is equivalent to Eq.~\eqref{eqHW2} when $U(n)=1$ for $\forall n$.
(The overall factor $1/2$ can be understood as the rescaling of lattice spacing $a \to 2a$.)
Although the Wilson fermion and the Kogut-Susskind fermion are different in general, they are equivalent in the one-dimensional, massless, and non-interacting case.
The fermion number operator is
\begin{equation}
 \hat{Q} =  \sum_{x}\bigg\{\hat{\psi}_e(x) ^\dagger\hat{\psi}_e(x) + \hat{\psi}_o(x) ^\dagger\hat{\psi}_o(x) - 1\bigg\},
\end{equation}
the naive chiral charge operator is
\begin{equation}
 \hat{Q}^5_{\rm naive} = i\sum_x 
 \bigg\{ 
 - \hat{\psi}_o^\dagger(x) \hat{\psi}_e(x) + \hat{\psi}_e^\dagger(x) \hat{\psi}_o(x)
 \bigg\},
\end{equation}
and the conserved chiral charge operator is
\begin{equation}
\begin{split}
 \hat{Q}^5 =& \frac{i}{2} \sum_x \bigg\{ -\hat{\psi}_o^\dagger(x)  \hat{\psi}_e(x) + \hat{\psi}_e^\dagger(x)  \hat{\psi}_o(x) \\
&- \hat{\psi}_o^\dagger(x+1) \hat{U}^\dagger(2x) \hat{U}^\dagger(2x-1) \hat{\psi}_e(x) \\
&+ \hat{\psi}_e^\dagger(x) \hat{U}(2x-1)\hat{U}(2x) \hat{\psi}_o(x+1)  \bigg\} . 
\label{eqQ5KS}
\end{split}
\end{equation}
The commutation relations are $[\hat{H}_f,\hat{Q}] =[\hat{H}_f,\hat{Q}^5] =0$ and $[\hat{H}_f,\hat{Q}^5_{\rm naive}] \neq 0$.

We assumed infinite space or periodic space for the above equations.
In quantum computing, the open boundary is often chosen because the gauge field can be eliminated by solving the Gauss law.
Unfortunately, the open boundary violates the conservation of the chiral charges \eqref{eqQ5W} and \eqref{eqQ5KS}, and this simplified formulation is not applicable.

\section{Chirality production}\label{sec3}

The time evolution of a quantum system is described by the evolution equation
\begin{equation}
 |\Psi(t) \rangle = \exp\left(-i\hat{H}t \right) |\Psi(0) \rangle .
\label{eqevo}
\end{equation}
When the system contains a gauge field and a fermion, the total Hamiltonian is 
\begin{equation}
 \hat{H} = \hat{H}_g + \hat{H}_f
\end{equation}
and the total state vector is
\begin{equation}
 |\Psi(t) \rangle = \prod_x |U(x)\rangle \otimes |\psi_1(x)\rangle \otimes |\psi_2(x)\rangle .
\end{equation}
When the Hamiltonian does not explicitly depend on time, the time evolution of the chiral charge is given by
\begin{equation}
 \frac{d}{dt} \langle \Psi(t) | \hat{Q}^5 | \Psi(t)\rangle = \langle \Psi(t) | i[\hat{H}_g,\hat{Q}^5] | \Psi(t)\rangle
 \label{eqdqdt}
\end{equation}
because of $[\hat{H}_f,\hat{Q}^5] =0$.
When the gauge field does not exist, $\hat{H}_g=0$, the chiral charge of a free fermion is conserved,
$\frac{d}{dt} \langle \Psi(t)|\hat{Q}^5|\Psi(t)\rangle = 0$.
When the gauge field exists, the link operator $\hat{U}(x)$ and its conjugate operator, i.e., the electric field operator $\hat{E}(x)$, satisfy the equal-time canonical commutation relation $[\hat{E}(x),\hat{U}(y)]=e\delta_{xy}\hat{U}(y)$.
Since the chiral charge operator contains the link operator and the gauge field Hamiltonian contains the electric field operator, they do not commute, $[\hat{H}_g,\hat{Q}^5] \neq 0$.
Therefore, the chiral charge conservation is anomalously violated, $\frac{d}{dt}  \langle \Psi(t)|\hat{Q}^5|\Psi(t)\rangle \neq 0$.
This non-conservation would be interpreted as the chiral anomaly in the continuum limit.
For example, in the U($1$) lattice gauge theory, 
\begin{equation}
 \frac{d}{dt} \langle \Psi(t) | \hat{Q}^5 | \Psi(t)\rangle = \langle \Psi(t) |\frac{e}{\pi}\int dx \hat{E}(x)| \Psi(t)\rangle
 \label{eqanomaly}
\end{equation}
is expected when the excess of the axial vector current is zero at boundaries \footnote{To be honest, the direct perturbative calculation of Eq.~\eqref{eqdqdt} reproduces only a half of the right-hand side of Eq.~\eqref{eqanomaly}. The mystery of this mismatch will be revealed somewhere in a future work.}.

\section{Quantum simulation}\label{sec4}

The time evolution mentioned above will hopefully be simulated by quantum computers.
The simulation of a continuous gauge group is of physical interest since the theory has a well-defined continuum limit, but the simulation of a discrete gauge group is better for exercise.
In this paper, we demonstrate the quantum simulation in the simplest case, i.e., the $Z_2$ lattice gauge theory with the Wilson fermion.
We used a noiseless emulator although the simulation is feasible on real devices.
For the state-of-the-art simulation of the one-dimensional $Z_2$ lattice gauge theory on a real device, see Ref.~\cite{Charles:2023zbl}.

We write down the Pauli gate representation of operators.
The gauge field Hamiltonian is
\begin{equation}
 \hat{H}_g = - \lambda \sum_x X_g(x)
\label{eqHg}
\end{equation}
with a real parameter $\lambda$, and the link operator is
\begin{equation}
 \hat{U}(x) = Z_g(x) .
\end{equation}
By the Jordan-Wigner transformation, the fermionic operators \eqref{eqHW2} to \eqref{eqQ5W2} are written as
\begin{align}
\begin{split}
 \hat{H}_f &= \sum_x \frac{1}{2} \big\{ X_1(x) X_2(x) + Y_1(x) Y_2(x) \\
&\quad - Z_g(x) X_1(x) X_2(x+1) \\
&\quad - Z_g(x) Y_1(x) Y_2(x+1) \big\}
\end{split}\\
 \hat{Q} &= \sum_x \frac{1}{2} \left\{ Z_1(x) + Z_2(x) \right\} \\
 \hat{Q}^5_{\rm naive} &= \sum_x \frac{1}{2} \left\{ Y_1(x) X_2(x) - X_1(x) Y_2(x) \right\} \\
\begin{split}
 \hat{Q}^5 &= \sum_x \frac{1}{4} \big\{ Y_1(x) X_2(x) - X_1(x) Y_2(x) \\
&\quad + Z_g(x) Y_1(x) X_2(x+1) \\
&\quad - Z_g(x) X_1(x) Y_2(x+1) \big\} .
\end{split}
\end{align}
The subscripts $g$, $1$, and $2$ stand for the gates acting on $|U(x)\rangle$, $|\psi_1(x)\rangle$, and $|\psi_2(x)\rangle$, respectively.
The quantum circuit for the evolution equation \eqref{eqevo} is the standard one. 
The evolution operator is decomposed into exponential operators by the Suzuki-Trotter formula with the step size $\delta t = 0.1$.
Since each exponential operator acts on up-to-three qubits, the circuit can be easily constructed.

\begin{figure*}[ht]
\begin{minipage}{0.45\textwidth}
\begin{center}
 \includegraphics[width=0.8\textwidth]{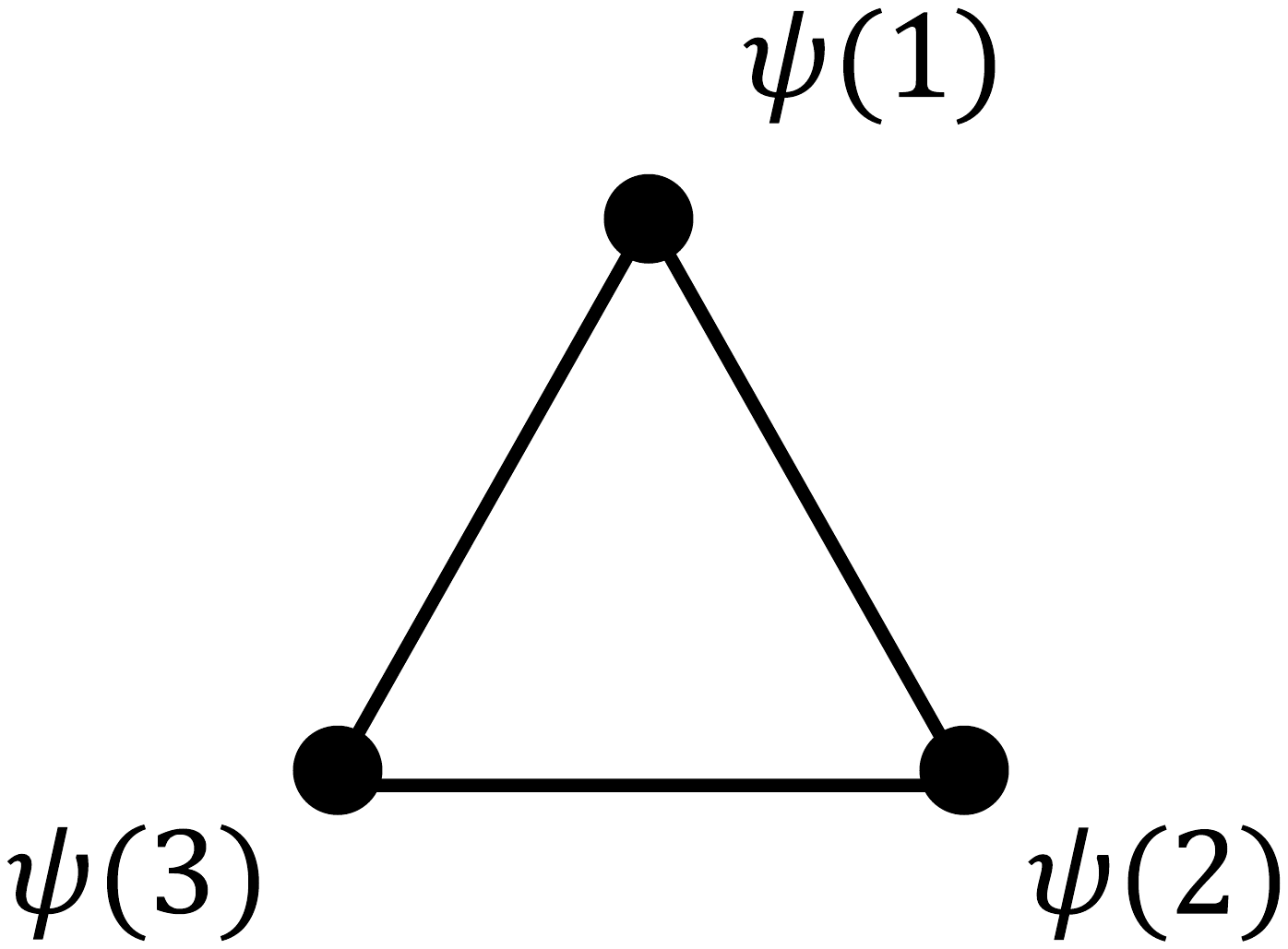}
\end{center}
\end{minipage}
\begin{minipage}{0.45\textwidth}
\begin{center}
 \includegraphics[width=1\textwidth]{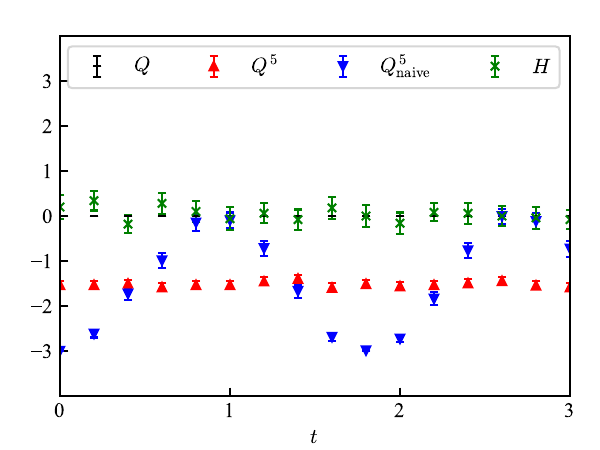}
\end{center}
\end{minipage}
\caption{
\label{fig1}
Time evolution of the free Wilson fermion.
The fermion number $\langle Q \rangle$, the conserved chiral charge $\langle Q^5 \rangle$, the naive chiral charge $\langle Q^5_{\rm naive} \rangle$, and the energy $\langle H \rangle$ are shown.
The simulation was done by a noiseless emulator and the error bar is a statistical error.
}
\end{figure*}

For a validity check of the simulation, we first study the non-interacting Wilson fermion.
Figure \ref{fig1} shows lattice geometry and the result.
The number of lattice sites is three and the boundary condition is periodic.
The state vector can be stored in six qubits.
The initial condition is chosen as an eigenstate of $\gamma^5$, s.t., $\hat{\psi}^\dagger(x)\gamma^5\hat{\psi}(x)|\psi_1(x)\rangle \otimes |\psi_2(x)\rangle=-|\psi_1(x)\rangle \otimes |\psi_2(x)\rangle$ for $\forall x$, although there is no special reason for this choice.
When there is no gauge interaction, there are three conserved quantities: the fermion number, the chiral charge, and the energy.
The naive chiral charge is not conserved due to the lattice discretization artifact, i.e., Eq.~\eqref{eqHQ5Wnaive}.
These expected behaviors are confirmed in Fig.~\ref{fig1}.

\begin{figure*}[ht]
\begin{minipage}{0.45\textwidth}
\begin{center}
 \includegraphics[width=0.8\textwidth]{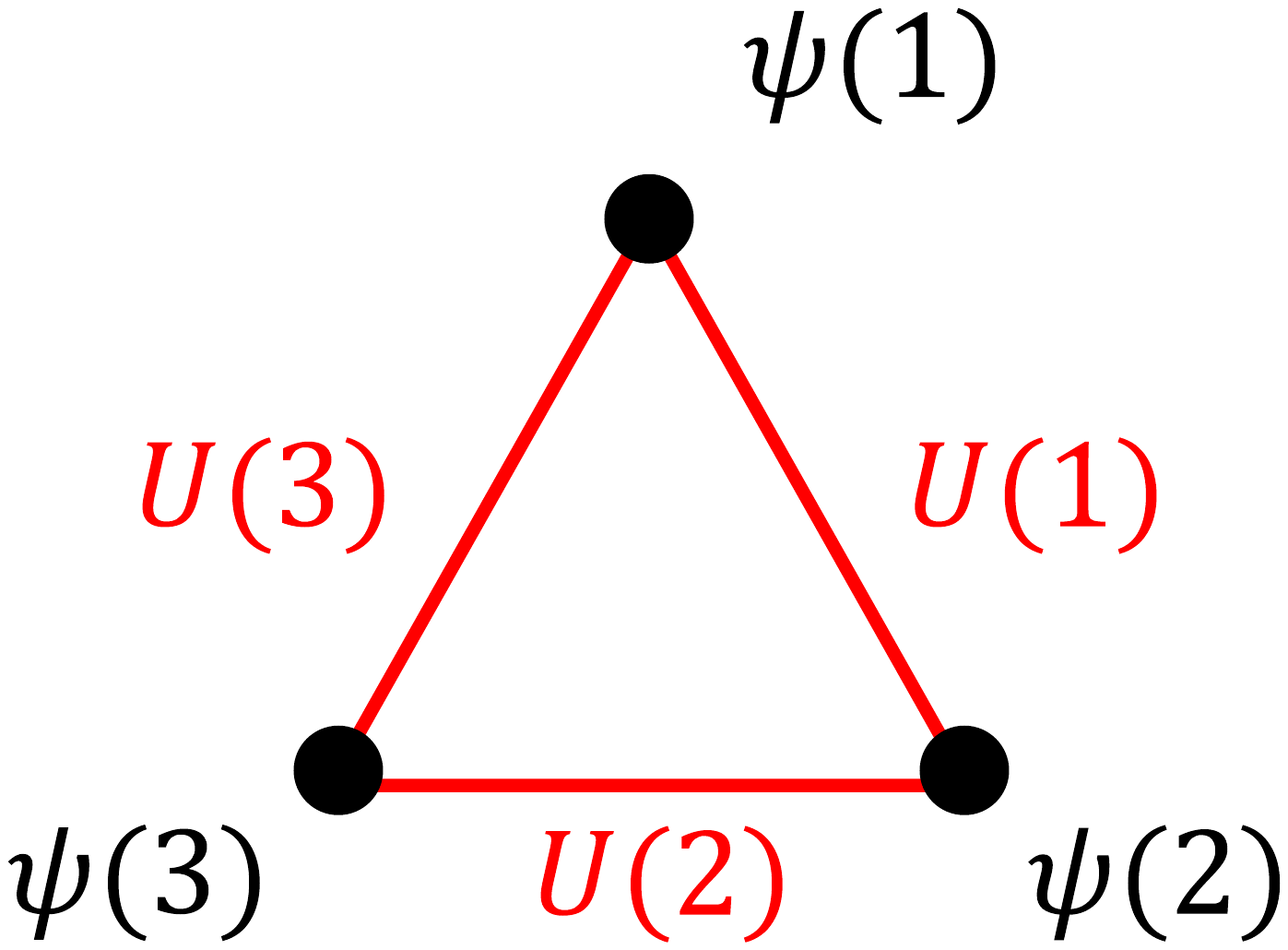}
\end{center}
\end{minipage}
\begin{minipage}{0.45\textwidth}
\begin{center}
 \includegraphics[width=1\textwidth]{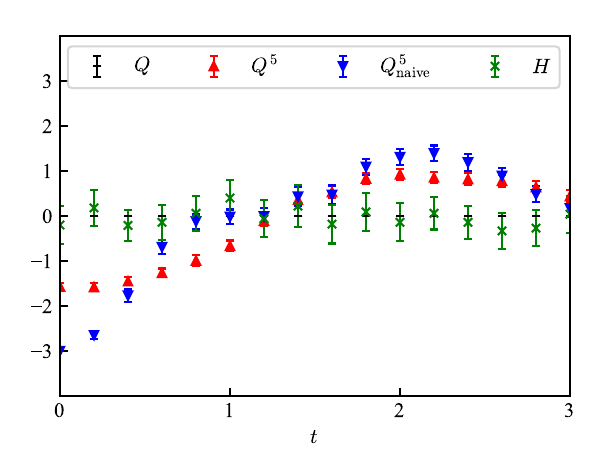}
\end{center}
\end{minipage}
\caption{
\label{fig2}
Time evolution of the Wilson fermion with the $Z_2$ gauge field.
The fermion number $\langle Q \rangle$, the conserved chiral charge $\langle Q^5 \rangle$, the naive chiral charge $\langle Q^5_{\rm naive} \rangle$, and the energy $\langle H \rangle$ are shown.
The simulation was done by a noiseless emulator and the error bar is a statistical error.
The gauge parameter is fixed at $\lambda=1$.
}
\end{figure*}

The interacting case is shown in Fig.~\ref{fig2}.
The $Z_2$ gauge fields exist on three links, so the state vector can be stored in nine qubits.
The initial condition of the gauge field is a trivial state, s.t., $\hat{U}(x)|U(x)\rangle=|U(x)\rangle$ for $\forall x$.
Other simulation setups are the same as in the non-interacting case.
The chiral charge is not conserved but generated by the gauge interaction.
(Even though the naive chiral charge shows a similar tendency to the conserved chiral charge, it is a distinguishable mixture of the physical time evolution and the lattice discretization artifact.)
The further interpretation of this chirality generation is nontrivial in the $Z_2$ lattice gauge theory.
The $Z_2$ lattice gauge theory does not have a continuum limit.
The chiral anomaly originates from ultraviolet divergence, so it appears only in a continuum limit.
It is correct to say that this chirality generation is due to the gauge interaction but subtle to say that the generation is due to the chiral anomaly.
Another subtlety is that the generation is not controlled by the external electric field.
In continuous gauge theories, the electric field is a continuous variable.
Positive external electric field increases the chiral charge and negative external electric field decreases the chiral charge.
In the $Z_2$ lattice gauge theory, there are only two eigenstates of the electric field: 0 and $\pi$ $({\rm mod} \ 2\pi)$.
There is no distinction between positive and negative electric fields.
Whether the chiral charge is increased or decreased, i.e., the sign of $\frac{d}{dt}\langle \hat{Q}^5 \rangle$, is determined by the initial condition, not by the external electric field.

The above analysis is applicable to the continuous gauge group if sufficient resources are given.
For example, $U(1)$ is approximated by $Z_N$ with large $N$.
The basic strategy is the same although the quantum circuit will be complicated and lengthy.
The simulation will reproduce the chirality generation by the chiral anomaly.
The chiral anomaly in one dimension is a simple but interesting subject for near-term quantum computation of lattice gauge theory.

\begin{acknowledgments}
The numerical calculations were performed with use of the QasmSimulator in IBMs open source SDK Qiskit~\cite{Qiskit}.
This work was supported by JSPS KAKENHI Grant No.~19K03841, 21H01007, and 21H01084.
\end{acknowledgments}

\bibliographystyle{apsrev4-2}
\bibliography{paper}

\end{document}